\documentclass[10pt,superscriptaddress,twocolumn,showpacs,pra,nofootinbib]{revtex4-1}
\usepackage{bm} 
\usepackage{graphicx} 
\usepackage{amsmath}
\usepackage{amsthm}
\usepackage{amssymb} 
\usepackage{comment} 
\usepackage{epstopdf} 
\usepackage{hyperref}
\usepackage{hypernat}
\usepackage{amsfonts}
\usepackage{color} 
\usepackage{subfigure}
\usepackage[english]{babel}
\usepackage{enumitem}
 \usepackage[bb=boondox]{mathalfa}

\newcommand{\bra}[1]{\langle #1 |}
\newcommand{\ket}[1]{| #1 \rangle}

\newcommand {\be}{\begin{equation}}
\newcommand {\ee}{\end{equation}}

\newcommand\s{{\text{stoq}}}

\newcommand{\C}{{\mathcal{C}}}
\newcommand{\ba}{\begin{eqnarray}}
\newcommand{\ea}{\end{eqnarray}}
\newcommand\tr{{\mbox{Tr\,}}}
\newcommand{\ignore}[1]{}

\newcommand{\bes} {\begin{subequations}}

\newcommand{\ees} {\end{subequations}}
\newcommand\D{{\text{c}}}
\newcommand\cl{{\text{c}}}
\newcommand\sgn{{\text{sgn}}}

\DeclareRobustCommand\openzero{\leavevmode{0\kern-.55em0}}
\mathchardef\minus="002D

\usepackage[margin=1in,nofoot]{geometry}

\newtheorem{definition}{Definition}

\newcommand{\beq}{\begin{eqnarray}}
\newcommand{\eeq}{\end{eqnarray}}

\newcommand{\Tr}{{\mathrm{Tr}}}
\newcommand{\e}{{e}}

\renewcommand{\Re}{\operatorname{Re}}

\newcommand {\bea}{\begin{eqnarray}}
\newcommand {\eea}{\end{eqnarray}}

\newcommand{\bwide}{\begin{widetext}}
\newcommand{\ewide}{\end{widetext}}

\begin{document}

\title{Elucidating the interplay between non-stoquasticity and the sign problem}
\author{Lalit Gupta}
\affiliation{Department of Physics and Astronomy and Center for Quantum Information Science \& Technology, University of Southern California, Los Angeles, California 90089, USA}
\author{Itay Hen}
\affiliation{Department of Physics and Astronomy and Center for Quantum Information Science \& Technology, University of Southern California, Los Angeles, California 90089, USA}
\affiliation{Information Sciences Institute, University of Southern California, Marina del Rey, California 90292, USA}
\email{itayhen@isi.edu}
\begin{abstract}
\noindent The sign problem is a key challenge in computational physics, encapsulating our inability to properly understand many important quantum many-body phenomena in physics, chemistry and the material sciences. Despite its centrality, the circumstances under which the problem arises or can be resolved as well as its interplay with the related notion of `non-stoquasticity' are often not very well understood. In this study, we make an attempt to elucidate the circumstances under which the sign problem emerges and to clear up some of the confusion surrounding this intricate computational phenomenon. To that aim, we make use of the recently introduced off-diagonal series expansion quantum Monte Carlo scheme with which we analyze in detail a number of examples that capture the essence of our results. 
\end{abstract}

\maketitle

\section{Introduction}

The `negative sign problem,' or simply the `sign problem,' is the single most important unresolved challenge in quantum many-body simulations, preventing physicists, chemists and material scientists alike from a true understanding of many of the most profound 
macroscopic quantum physical phenomena of Nature---in areas as diverse as material design and high temperature superconductivity through the physics of neutron stars to lattice quantum chromodynamics and more~\cite{Wiese-PRL-05,marvianLidarHen,klassenMarvian,signProbSandvik}. 

The sign problem slows down quantum Monte Carlo (QMC) algorithms~\cite{Alet,PhysRevB.93.054408,PhysRevB.89.134422}, which are in many cases the only tool available to us for studying large quantum many-body systems, to the point where these schemes become practically useless. QMC algorithms allow us to evaluate thermal averages of physical observables by sampling the configuration space of the model in question via the decomposition of the partition function into a sum of efficiently computable terms, which are in turn interpreted as probabilities in a Markov process~\cite{Landau:2005:GMC:1051461,newman}.
The sign problem emerges whenever a decomposition of the quantum partition function into a sum of non-negative terms is not known, in which case the speedups that normally accompany importance sampling are lost or considerably diminished. 

Despite the centrality of the sign problem to the understanding of many important physical phenomena, certain aspects of the sign problem, such as the circumstances under which the problem emerges, the practical meaning of resolving the problem and its interplay with the related notion of `non-stoquasticity' that has gained traction within the physics and computer sciences communities in its own right in recent years~\cite{Bravyi:QIC08,Bravyi:2014bf,marvianLidarHen}, are often not very well understood. 

In an effort to resolve some of the misconceptions surrounding this important computational phenomenon, in this study, we examine in detail the emergence of the sign problem in QMC and clarify certain aspects of it. To that aim, we make use of the recently introduced off-diagonal series expansion quantum Monte Carlo scheme~\cite{ODE,ODE2,pmr}, a method that builds on a power series expansion of the quantum partition function in its off-diagonal terms and is both parameter-free and Trotter error-free. We illustrate our key results by analyzing several toy examples that capture the essence of these. 

\section{The QMC sign problem}
\subsection{Basic definitions}
We begin our discussion by providing a practical definition to the sign problem in the context of QMC (thereby also acknowledging the existence of other definitions~\cite{Wiese-PRL-05}). This definition will then allow us to discuss in more detail the circumstances under which the problem emerges and also the practical meaning of curing it.  

\begin{definition}
An $N$-particle quantum many-body model given by a Hamiltonian $H_N$ will be said to possess a sign problem if there is no known decomposition of its partition function $Z=\Tr[\e^{-\beta H_N}]$ into a sum of efficiently computable (in $\beta$ and in $N$) positive-valued terms. 
\end{definition}

It is important to note that our definition of the sign problem has to do with the existence of a decomposition of the partition function $Z=\Tr[\e^{-\beta H}]$ into a sum of efficiently computable positive weights $Z=\sum_\C W_\C$ (where each weight $W_\C$ corresponds to a configuration $\mathcal{C}$), rather than with the computational effort required to sample these terms or to accurately evaluate thermal averages of physical observables. 

The above definition also implies that `curing' or resolving the sign problem for a model $H$, i.e., finding a decomposition of its partition function into efficiently computable positive weights, implies nothing about the computational efficiency with which the QMC algorithm samples the configuration space. Expressed differently, the absence of a sign problem does not guarantee efficient QMC convergence. An example that illustrates this fact best is that of classical spin glasses (or Ising models). These are classical systems. Their partition functions can be written as sums of strictly positive Boltzmann weights and as such they trivially have no sign problem (by the same token, one may also consider quantum spin glasses by augmenting the Ising model with a small transverse field). Nonetheless these systems are known to equilibrate in the worst case in exponential time~\cite{Barahona1982}. 

\subsection{Determining the severity of the sign problem}

The reason the sign problem is indeed a serious impediment to QMC algorithms is that the computational efficiency of QMC hinges on the algorithm's ability to efficiently sample the configurations 
$\mathcal{C}$ as obtained from the partition function decomposition according to their importance, or relative weight,
$p_\C=W_\C/Z$.
A necessary but not sufficient condition for proper importance-based sampling (or importance sampling for short) is that all weights $W_\C$ are positive (or nonnegative), i.e., that $p_\mathcal{C}$ can be interpreted as a bona fide probability distribution over configurations. 
When this happens, the thermal average $\langle A \rangle$ of any physical observable $A$ can be written as
\beq \label{eq:expectForm}
\langle A \rangle  = \frac{\Tr \left[A \e^{-\beta H}\right]}{Z} = \frac{\sum_\C A_\C W_\C}{\sum_\C W_\C} = \sum_\C A_\C p_\C \,.
\eeq
Since an explicit summation over all terms in the sum above is in general prohibitive due to the generally exponential number of terms in the sum, QMC importance sampling estimates this sum using a Monte Carlo estimator defined as 
\begin{eqnarray}
{\langle \tilde{A} \rangle}_p \approx \frac{1}{N_\text{s}} \sum_{i=1}^{N_\text{s}} A_{\mathcal{C}_i}\,,
\end{eqnarray}
where the configurations $\C_i$ (of which there are $N_\text{s}$) are randomly sampled in proportion to their probability $p_{\mathcal{C}_i}$~\cite{MonteCarlo_Ambegaokar}. If the number of `important configurations' is relatively small, then estimation via importance sampling will converge much faster than with straightforward unbiased sampling of the various terms (or an actual summation over all terms) and often times exponentially more so.

Importance sampling is however not always possible. Whenever a subset of configurations is assigned negative weights, one cannot draw samples according to $p_\C$ (as $p_\C$ is no longer a probability distribution). A common workaround for the appearance of negative weights (hence the terminology `sign problem') is to draw samples from a different distribution, $\tilde{p}_\C$, that is nonnegative everywhere ~\cite{Rubinstein1981,Wiese-PRL-05}. Then, the estimator of $A$ becomes
\begin{eqnarray}
{\langle \tilde{A} \rangle}_{\tilde{p}} \approx \frac{1}{N_\text{s}} \sum_{i=1}^{N_\text{s}} A_{\mathcal{C}_i} \frac{p_{\mathcal{C}_i}}{\tilde{p}_{\mathcal{C}_i}}
\end{eqnarray}
A common choice for $\tilde{p}_\C$ is 
\beq
\tilde{p}_\C= \frac{|W_\C|}{\sum_{\C'} |W_{\C'}|}\,,
\eeq
i.e., weights that are proportional to the absolute values of the original weights. In the next section, we will see that the sum $\sum_\C |W_\C|$ may be viewed as the decomposition of the partition function of a related but not equivalent sign-problem-free model. 

With the above choice, thermal averages can be written as 
\beq \label{eqt:TEV}
\langle A\rangle = \frac{\sum_\C A_\C  \text{sgn}(W_\C) |W_\C|}{\sum_\C \text{sgn}(W_\C) |W_\C|} =\frac{\langle A \,\text{sgn}(W)\rangle_{|W|}}{\langle \text{sgn}(W) \rangle_{|W|}}  \,.
\eeq 
The subscript ${|W|}$ is added as a reminder that the  weights used in the Markov process are the absolute values of the original `true' weights. 

For models that do not have a sign problem, all weights are positive and $\langle \text{sgn}(W) \rangle_{|W|}=1$. For sign-problematic systems, negative weights are equally dominant and we have $\langle \text{sgn}(W) \rangle_{|W|}\approx 0$. In this case thermal averages of physical observables will fluctuate rapidly, resulting in extremely large error bars and will require an exponentially large number of measurements~\cite{Wiese-PRL-05,signProbODE}. 

An appropriate figure of merit for how adverse the sign problem is for a given QMC algorithm can therefore be given by the `weighted sign' $\langle \text{sgn}(W) \rangle_{|W|}$ (or $\langle \text{sgn}\rangle$ for short) which can be explicitly written as 
\beq\label{eq:sgnDef}
\langle \text{sgn}\rangle=\frac{\sum_\C W_\C}{\sum_\C |W_\C|}\,.
\eeq 

\subsection{Curing the sign problem}

Often times, even though a positive-valued decomposition is not known for the partition function of $H$, it may be known for that of a `rotated' model $\tilde{H}= U H U^\dagger$ where $U$ is a (usually local) unitary transformation. 
Since 
\bea
\tilde{Z}&=&\Tr \left[ \e^{-\beta \tilde{H}} \right]=\Tr \left[ \e^{-\beta U H U^\dagger} \right] \nonumber\\
&=&\Tr \left[ U \e^{-\beta H } U^\dagger\right] = \Tr \left[ \e^{-\beta H }\right] =Z \,,
\eea
the two models $\tilde{H}$ and $H$ are physically equivalent via a similarity transformation. 

The existence of the sign problem therefore depends on the basis in which the hamiltonian $H$ is represented in the QMC algorithm (but not only). 
An extreme example is the classical Ising Hamiltonian $H=\sum_{i<j} J_{ij} Z_i Z_j$ with $J_{ij}\in \{-1,1\}$ which is diagonal in the computational basis and thus has no sign problem (here $Z_i$ and $X_i$ denote the Pauli-$z$ and Pauli-$x$ operators acting on the $i$-th spin, respectively). 
In the rotated basis in which $Z_i \leftrightarrow X_i$ the same Hamiltonian will be written as $H=\sum_{i<j} J_{ij} X_i X_j$ and will possess a sign problem~\cite{marvianLidarHen}. Thus, often times curing the sign problem boils down to finding a transformation $U$ to the Hamiltonian 
 such that the decomposition of the partition function of the rotated Hamiltonian $\tilde{H}$ to positive-valued  terms is known~\cite{marvianLidarHen,Klassen2019twolocalqubit,klassenMarvian,eisertSignProb}. It should be noted nonetheless that since in QMC some physical observables are more easily accessible than others depending on whether or not they are diagonal, curing the sign problem via rotation is not always a practical resolution as observables in the rotated frame may be inaccessible as compared to those in the original frame. 
 
Rotating the Hamiltonian is not the only way to resolve the sign problem. Another method that is sometimes applicable, is using re-summation techniques wherein one groups together positive and negative QMC weights into strictly positive `grouped weights' (see, e.g., Refs.~\cite{resum1,resum2,signProbODE}). Here, no rotation takes place. 

As was already noted and is worth emphasizing again, finding a decomposition of the partition function into positive-valued efficiently computable terms cures the sign problem but promises nothing about the convergence time of the QMC simulation of the cured model. Expressed differently, freeing a model from its sign problem does not guarantee the efficient equilibration of the QMC Markov process, which depends on the details of the model in question and those of the QMC algorithm itself. 

\section{Off-diagonal series expansion of the quantum partition function\label{sec:par}}

As a vehicle for illustrating the main observations pertaining to the sign problem that we analyze in this work, we will be using for concreteness one particular flavor of QMC, namely, the off-diagonal expansion (ODE) QMC, first introduced in Refs.~\cite{ODE,ODE2}.  As we shall illustrate, the fact that ODE is both Trotter-error-free and parameter-free will allow us to distill certain aspects of the sign problem that are arguably less apparent when examined with other QMC schemes. We proceed with an overview of the technique. 

\subsection{ODE: an overview} \label{sec:ODE}
Within the off-diagonal expansion one considers many-body systems whose Hamiltonians we cast as the sum
\beq \label{eq:basic}
H=\sum_{j=0}^M \tilde{P}_{j} =\sum_{j=0}^M D_j P_j  \,,
\eeq
where $\{ \tilde{P}_j\}$ is a set of $M+1$ distinct generalized permutation matrices~\cite{gpm}, i.e., matrices with precisely one nonzero element in each row and each column (this condition can be relaxed to allow for zero rows and columns). 
Each operator $\tilde{P}_j$ can be written, without loss of generality, as $\tilde{P}_j=D_j P_j$ where $D_j$ is a diagonal matrix
and $P_j$ is a  permutation matrix with no fixed points (equivalently, no nonzero diagonal elements) except for the identity matrix $P_0=\mathbb{1}$. We will refer to the basis in which the operators $\{D_j\}$ are diagonal as the computational basis and denote its states by $\{ |z\rangle \}$. We will call the diagonal matrix $D_0$ the `classical Hamiltonian' and will sometimes denote it by $H_{\cl}$. The reader is referred to Ref.~\cite{pmr} for additional details. 

The $\{D_j P_j \}$  off-diagonal operators (in the computational basis) give the system its  `quantum dimension'.  Each term $D_j P_j $ obeys 
$D_j P_j | z \rangle = d_{z'}^{j} | z' \rangle$ where $d_{z'}^{j}$ is a possibly complex-valued coefficient and $|z'\rangle \neq |z\rangle$ is a basis state.
While the above formulation may appear restrictive, any finite-dimensional matrix can be written in the form of Eq.~\eqref{eq:basic}. 

\begin{figure}[htp]
\includegraphics[width=.48\textwidth]{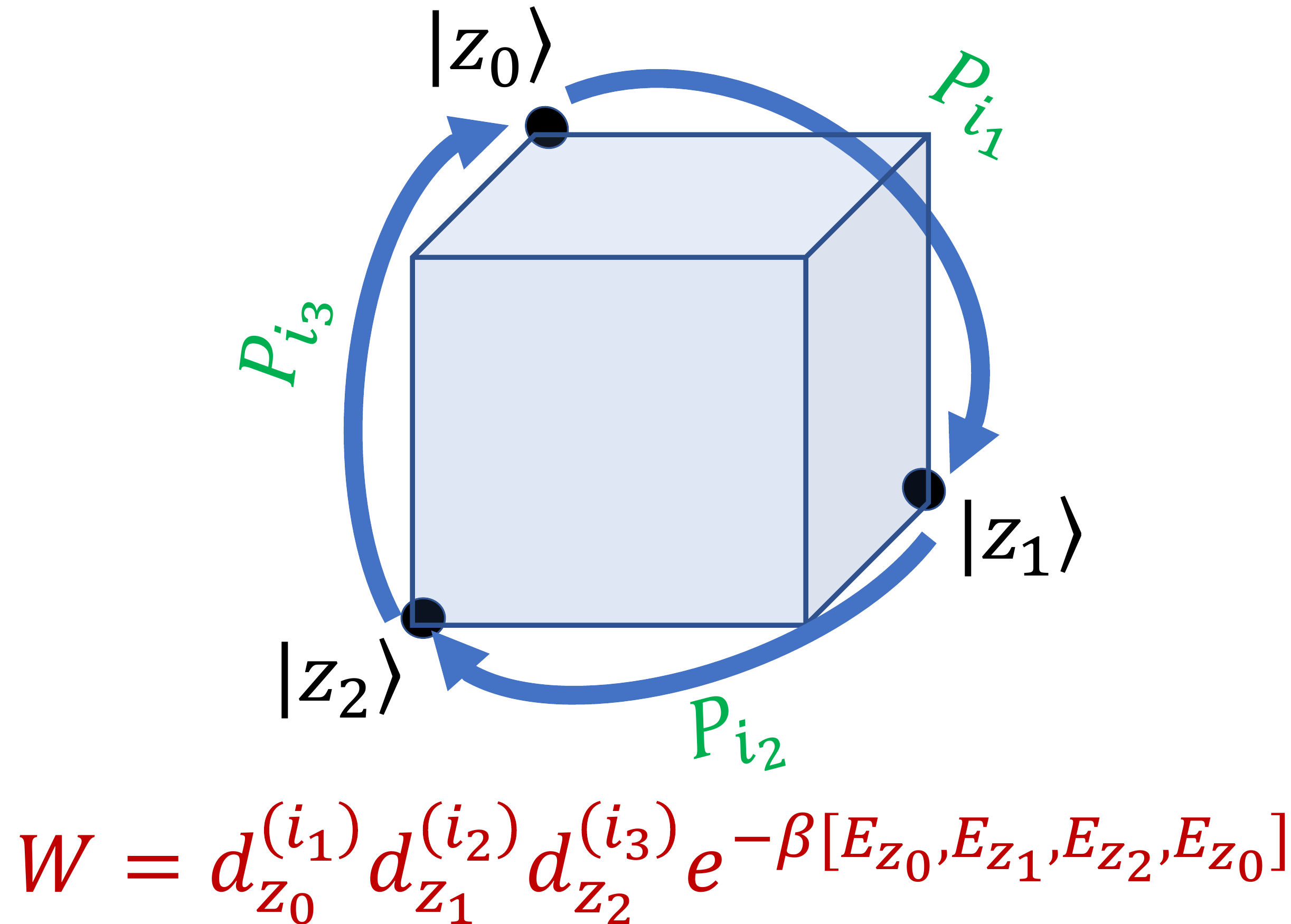}
\caption{\label{fig:hyper} Diagrammatic representation of a generalized Boltzmann weight, or a GBW, calculated from the classical energies $E_{z_j}$ of the classical states $|z_j\rangle$, which form a closed path, or a cycle, in the hypercube of basis states. The path is determined by the action of the permutation operators of the configuration, represented by $S_{{\bf{i}}_q}=P_{i_3} P_{i_2} P_{i_1}$, on the initial basis state $|z_0\rangle$. Cycles close if and only if the sequence of permutation operators evaluates to the identity operation. 
}
\end{figure}

Given the above formulation for the Hamiltonian, the off-diagonal series expansion of the partition function $Z=\tr\left[ \e^{-\beta H} \right]$ proceeds as follows.
Expanding the exponential in a Taylor series  in the inverse-temperature $\beta$, $Z$ can be written as a triple sum over all basis states $|z\rangle$, the expansion order $q$ which ranges from 0 to infinity and the (unevaluated) products \hbox{$S_{\bf{i}_q} = P_{i_q} \ldots P_{i_2} P_{i_1}$} of $q$ off-diagonal operators. Here we have used the multiple index ${\bf i}_q = (i_1,\ldots,i_q)$ where each individual index $i_j$ ranges from $1$ to $M$. In this notation, the empty sequence $S_{\bf{i}_0}$ corresponds to the identity operation. 
After some algebra, the partition function attains the form
 \beq \label{eq:z1}
Z =\sum_{\{z\}} \sum_{q=0}^{\infty}  \sum_{\{S_{{\bf{i}}_q}\}}  D_{(z,S_{{\bf{i}}_q})} \bra{z} S_{{\bf{i}}_q} \ket{z}   \e^{-\beta [E_{z_0},\ldots,E_{z_q}]} \,,\nonumber\\
\eeq 
where $\{S_{{\bf{i}}_q}\}$ is the set of all (unevaluated) products \hbox{$P_{i_q} \ldots P_{i_2} P_{i_1}$} of size $q$ and the term $e^{-\beta[E_{z_0},\ldots,E_{z_q}]}$ is the \emph{exponent of divided differences} (see the Appendix) over the multiset of classical energies $[E_{z_0},\ldots E_{z_q}]$ ~\cite{dd:67,deboor:05}.
The energies $\left\{E_{z_i}=\langle z_i | H_\D|z_i\rangle \right\}$ are the classical energies of the states $|z_0\rangle, \ldots, |z_q\rangle$ obtained from the action of the ordered $P_j$ operators in the sequence $S_{{\bf{i}}_q}$ on $|z_0\rangle$, then on $|z_1\rangle$, and so forth.
Explicitly, $|z_0\rangle=|z\rangle, P_{i_1}|z_0\rangle=|z_1\rangle, P_{i_2}|z_1\rangle=|z_2\rangle$, etc. The sequence of basis states $\{|z_i\rangle \}$ may be viewed as a `path' in the hypercube of basis states (see Fig.~\ref{fig:hyper}).\footnote{Note that \hbox{$|z_j\rangle=P_{i_j} \ldots P_{i_2} P_{i_1}|z\rangle$} should in principle have been denoted $|z_{(i_1,\ldots,i_j)}\rangle$. We are using a simplified notation so as not to overburden the notation.} 

Additionally, we have denoted 
\beq
D_{(z,S_{{\bf{i}}_q})}=\prod_{j=1}^q d^{(i_j)}_{z_j}\,,
\eeq
 where 
\beq \label{eq:hs}
d^{(i_j)}_{z_j} = \langle z_j | D_{i_j}|z_j\rangle \,,
\eeq
can be considered as the `hopping strength' of $P_{i_j}$ with respect to $|z_j\rangle$. 
Note that while the partition function is positive and real-valued, the $d^{(i_j)}_{z_j}$ elements do not necessarily have to be so. 

Having derived the expansion Eq.~(\ref{eq:z1}) for any Hamiltonian cast in the form Eq.~(\ref{eq:basic}), we are now in a position to interpret the partition function expansion as a sum of weights, i.e., $Z = \sum_{{\cal C}} W_{{\cal C}}$, where the set of configurations $\{{\cal C}\}$ is all the distinct pairs $\{ |z\rangle, S_{{\bf{i}}_q} \}$.  Since $\langle z| S_{{\bf{i}}_q} |z\rangle$ is either zero or one, we can write $W_{{\cal C}}$ as 
\beq \label{eq:gbw}
W_{{\cal C}}=   D_{(z,S_{{\bf{i}}_q})}  \e^{-\beta [E_{z_0},\ldots,E_{z_q}]}\,,
\eeq
we refer to it as  a `generalized Boltzmann weight' (or, a GBW).  It can be shown~\cite{ODE} that $(-1)^q \e^{-\beta [E_{z_0},\ldots,E_{z_q}]}$ is strictly positive. 

\subsection{The sign problem within ODE}
As mentioned above, the weights $W_{\cal C}$ will in general be complex-valued, despite the partition function being real (and positive). Since however for every configuration $\mathcal{C}=\{ |z\rangle, S_{{\bf{i}}_q}\}$ there is a conjugate configuration $\bar\C=\{ |z\rangle, S^{\dagger}_{{\bf i}_q}\}$\footnote{For $S_{{\bf{i}}_q} =P_{i_q} \ldots  P_{i_2} P_{i_1}$, the conjugate sequence is simply $S^\dagger_{{\bf i}_q} = P_{i_1}^{-1} P_{i_2}^{-1} \ldots P_{i_q}^{-1}$.} that produces the conjugate weight $W_{\bar\C}=\bar{W}_\C$, the imaginary contributions cancel out.  Expressed differently, the imaginary portions of complex-valued weights do not contribute to the partition function and may be disregarded altogether. We may therefore redefine \hbox{$D_{(z,S_{{\bf{i}}_q})}=\Re\left[\prod_{j=1}^q d^{(i_j)}_{z_j}\right]$}, obtaining strictly real-valued weights. 

Before we move on, we note that since $\langle z| S_{{\bf{i}}_q} |z\rangle$ evaluates either to 1 or to zero, the expansion can be more succinctly rewritten as
 \beq\label{eq:z}
Z  =\sum_{\bra{z} S_{{\bf{i}}_q} \ket{z}=1}  D_{(z,S_{{\bf{i}}_q})}   \e^{-\beta [E_{z_0},\ldots,E_{z_q}]} \,,
\eeq
i.e., as a sum over all closed paths on the hypercube of basis states~\cite{ODE,ODE2,signProbODE}.
Moreover, since the permutation matrices with the exception of $P_0$ have no fixed points, the condition $\langle z| S_{{\bf{i}}_q} |z\rangle=1$ implies $S_{{\bf{i}}_q}=\mathbb{1}$, i.e., $S_{{\bf{i}}_q}$ must evaluate to the identity element $P_0$ (note that the identity element itself does not appear explicitly in the sequences $S_{{\bf{i}}_q}$).

\section{\label{sec:ns}Non-stoquasticity and the emergence of the sign problem}

An attractive property of the formalism introduced above is that it allows us to identify the emergence of the sign problem in QMC via inspection of the weights $W_{\cal C}$, thereby making the connection between the emergence of the sign problem and the notion of non-stoquasticity, which has garnered increasing attention with the advent of quantum computers in recent years, more apparent~\cite{Bravyi:QIC08,Bravyi:2014bf,marvianLidarHen,klassenMarvian}. 

As already mentioned, to interpret the real-valued weight terms $W_\C$ as actual weights (equivalently, un-normalized probabilities), these must be nonnegative as it is the presence of negative weights that marks the onset of the infamous sign problem. In ODE, a weight is positive iff 
\beq \label{eq:PosCond}
(-1)^q D_{(z,S_{{\bf{i}}_q})}=\Re\left[\prod_{j=1}^q (-d^{(i_j)}_{z_j})\right]
\eeq
is positive, that is, a QMC algorithm will encounter a sign problem, equivalently a negative weight, during a simulation if and only if \emph{there exists a closed path on the hypercube of basis states along which $\Re\left[\prod_{j=1}^q (-d^{(i_j)}_{z_j})\right]<0$}. It should thus be clear that it is not the sign of off-diagonal entries that creates the sign problem, but rather the cumulative phase of closed paths in the hypercube of basis states that determines its occurrence. 

We can use the above observation to identify several sign-problem-free classes of models. A special class of models where the sign problem does not emerge, i.e., where $\Re\left[\prod_{j=1}^q (-d^{(i_j)}_{z_j})\right]\geq 0$ for all configurations, is that of `stoquastic' Hamiltonians~\cite{Bravyi:QIC08,Bravyi:2014bf,marvianLidarHen}, which we define as follows.

\begin{definition}
A Hamiltonian will be called (explicitly)\footnote{Other definitions of stoquatcity account for the fact that some Hamiltonians may be easily cast in stoquastic form by applying a suitable local unitary transformation ot the Hamiltonian~\cite{Bravyi:QIC08,Bravyi:2014bf,marvianLidarHen}.} stoquastic with respect to a basis $\mathcal{B}$ if the Hamiltonian matrix representation in that basis has only non-positive off-diagonal  elements.  
\end{definition}

For stoquastic Hamiltonians all $d^{(i_j)}_{z_j}$ are negative (or zero). 
In this case, all paths trivially yield positive-valued weights. 

The existence of positive off-diagonal terms does not however imply a sign problem for QMC.
An example of a sign-problem-free family of models is the transverse-field Ising Hamiltonian 
\beq\label{eq:Hising}
H =\sum_{i,j} J_{ij} Z_i Z_j +  \sum_j h_j Z_j + \Gamma \sum_j X_j  \, .
\eeq
for $\Gamma > 0$.  A slightly less trivial example is the two-body model
\beq\label{eq:Hising2}
H =\sum_{i,j} J_{ij} Z_i Z_j + \Gamma \sum_{\langle i,j\rangle} X_i X_j  \, ,
\eeq
provided that the underlying connectivity of the two-body $X$ terms is bi-partite. In this case, all nonzero weights come from paths  on the hypercube of basis states consisting of only even number of steps (or, even $q$) leading to only positive-valued weights despite the existence of off-diagonal elements with the wrong sign. 

We note though that both models above can be cast in explicitly stoquastic form by applying simple local rotations. The transverse-field Ising model, Eq.~(\ref{eq:Hising}) can be made stoquastic by the transformation $X_i \to -X_i$. For the model of Eq.~(\ref{eq:Hising2}) a similar rotation performed only on one of the two partitions of the lattice will flip the sign of the off-diagonal terms\footnote{Less trivial examples for non-stoquastic yet sign-problem-free models do exist however they are somewhat less informative. In fact it can be shown that models with a sign problem that is curable by local rotations are easy to manufacture but are nonetheless generally hard to cure~\cite{marvianLidarHen}}.


As discussed before, a way to deal with the existence of negative-valued weights is to sample configurations according to their absolute values. It is interesting to note that the absolute-valued weights $|W_\C|$ are precisely those belonging to the related but not-physically-equivalent `stoquasticized' model $H_\s$ whose diagonal operators 
$D_j$ for $j>0$ are related to those of the sign-problematic one by the relation 
\beq 
D_j\to -|D_j| \,.
\eeq 
Using the above notation, one can see that $\langle \text{sgn}\rangle$ can be written as:
\beq\label{eq:sgnDef1}
\langle \text{sgn}\rangle=\frac{\Tr\left[\e^{-\beta H}\right]}{\Tr\left[\e^{-\beta H_\s}\right]}
=\frac{\Tr\left[S\e^{-\beta H_\s}\right]}{\Tr\left[\e^{-\beta H_\s}\right]} = \langle S\rangle_{|W|}\,,\nonumber\\
\eeq 
where we have defined $S=\e^{-\beta H}\e^{\beta H_\s}$ and the severity of the sign problem is the thermal average of $S$ with respect to the stoquasticized model $H_\s$ whose weights are $|W_\C|$.

\section{Illustrative examples}

Having defined the sign problem in the context of ODE, we are now in a position to derive several observations that illustrate some of the properties of the sign problem. The goal here is to settle certain ambiguities that have been attached to its nature over the years. 

\subsection{A single qubit cannot have a sign problem} 

We begin by showing that a single spin-$1/2$ Hamiltonian cannot possess a sign problem regardless of the basis in which it is represented. Since any two dimensional hermitian matrix can be written as a linear combination of the Pauli matrices and the identity, the Hamiltonian of a single spin-$1/2$ particle can most generally be written as
\beq
H=\alpha_0 \mathbb{1} +\alpha_1 X + \alpha_2 Y + \alpha_3 Z \ ,
\eeq
where $X, Y$ and $Z$ are the matrix representations of the usual Pauli operators in the  basis that diagonalizes the Pauli-$z$ operator and $\alpha_0,\ldots,\alpha_3$ are real parameters.  In permutation-matrix representation, the Hamiltonian becomes
\beq\label{eq:single_spin}
H=D_0 P_0+ D_1 P_1
\eeq
with $P_0 =  \mathbb{1}$, $P_1 = X$, $D_0=H_\cl=\alpha_0 \mathbb{1}+ \alpha_3 Z$ and $D_1=\alpha_1 \mathbb{1}-i \alpha_2 Z$.

Since in this example the $S_{{\bf{i}}_q}$ are sequences consisting of only one type of non-identity permutation matrices, namely $P_1=X$, the expansion order $q$ must be even for $S_{{\bf{i}}_q}$ to evaluate to the identity element. This in turn results in 
$$\left[\prod_{j=1}^q (-d^{(i_j)}_{z_j})\right] = {(\alpha^2_1+\alpha^2_2)}^{q/2}$$
 being strictly nonnegative. We thus find that any single-qubit Hamiltonian is necessarily also sign-problem-free. The same is however not true for a single qutrit in which case a sign problem may arise. We discuss the qutrit case next. 

\subsection{\label{sec:qutrit}A spin-one particle can have a sign problem} 
We next show that unlike a spin $1/2$ particle, a single qutrit, or a spin-one particle, can possess a sign problem (in a given basis). In fact, it is the lowest-dimensional model that is capable of exhibiting the problem. 

The most general Hamiltonian for a single qutrit can be written as $H=D_0 \mathbb{1}+D_1 P_1+D_2 P_2$
where
\[
\mathbb{1}= 
\begin{bmatrix}
    1  &  0 & 0      \\
    0 &  1 &  0	 	 \\
     0 &  0 &  1  
\end{bmatrix}\;,
P_1 = 
\begin{bmatrix}
  0  &  0 & 1      \\
    1 &  0 &  0	 	 \\
     0 &  1 &  0     
\end{bmatrix}\;,
P_2 = 
\begin{bmatrix}
  0  &  1 & 0      \\
    0 &  0 &  1	 	 \\
     1 &  0 &  0     
\end{bmatrix} \, , 
\]
and $D_2 = D_1^\ast$, a condition required by hermiticity of the Hamiltonian. The operators obey 
$P_1 P_2=\mathbb{1}$ and $P_1^{\dagger}=P_1^2=P_2$.

\begin{figure}[htp]
\includegraphics[width=0.45\textwidth]{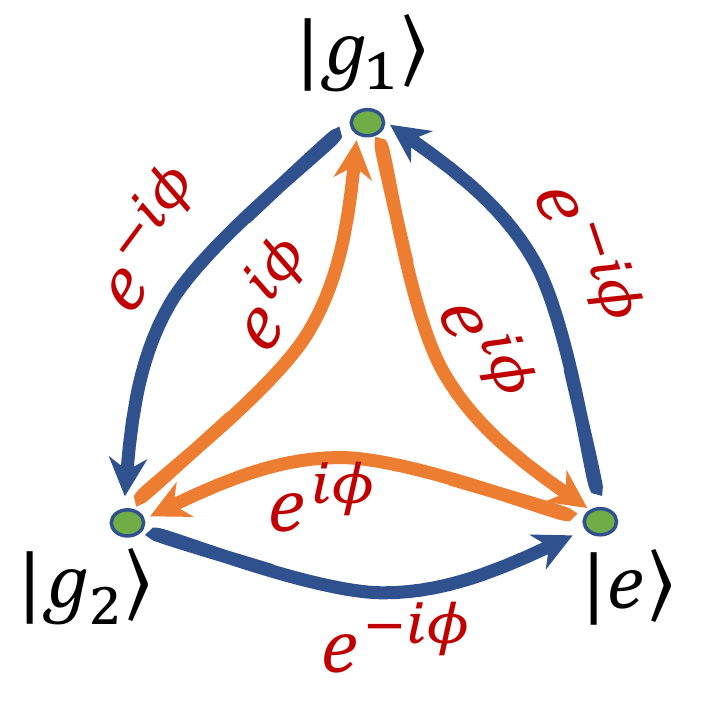}
\small
\caption{\label{fig:qutritPaths} Action of $D_1 P_1$ (orange arrows) and $D_2 P_2$ (blue arrows) on the three classical basis states $|g_1\rangle$, $|g_2\rangle$ and $|e\rangle$. The phases accompanying the permutations are also indicated.} 
\end{figure}

Here, we will consider the particular case
\beq\label{eq:qutritHam}
H= \e^{i \phi} P_1 + \e^{-i \phi} P_2 +J \text{diag}\{0,1,0\}\,.
\eeq
The diagonal part of the Hamiltonian is \hbox{$D_0=J \text{diag}\{0,1,0\}$} and has two zero-energy ground states $|g_1\rangle=\{1,0,0\}$ and $|g_2\rangle=\{0,0,1\}$ and a single excited state $|e\rangle=\{0,1,0\}$ with classical energy $J$ (we will be assuming $J>0$). 
The off-diagonal operators $D_{1/2}=\e^{\pm i \phi} \mathbb{1}$ are pure phases. The action of $D_1 P_1$ and $D_2 P_2$ on the three classical states is depicted in Fig.~\ref{fig:qutritPaths}. Needless to say, for such a small system, the Hamiltonian is easily diagonalizable and may be readily represented in the diagonilizing basis for arbitrary values of $\phi$ and $J$. In the diagonilizing basis there is of course no sign problem.


As noted earlier, an ODE configuration $\mathcal{C}$ consists of a basis state $|z\rangle$ and a product \hbox{$S_{\bf{i}_q}={V}_{i_1} \cdot {V}_{i_2} \cdots {V}_{i_q}$} of off-diagonal operators. In the qutrit model \hbox{$|z\rangle \in \{ |g_1\rangle,|g_2\rangle,|e\rangle\}$} and \hbox{$V_{i_j} \in \{ P_1, P_2\}$}.
A $\{|z\rangle, S_{\bf{i}_q}\}$ pair induces a sequence of classical states $|z_i\rangle$ generated by the action of the off-diagonal operators on the basis state $|z\rangle$. The sequence of basis states $\{|z_i\rangle \}$ may be viewed as a `path' in the hypercube of basis states. For a weight to have a nonzero value, the path must be a closed one, namely, $|z\rangle=|z_0\rangle=|z_q\rangle$. 
We present in Table~\ref{tab:paths} the shortest sequences of operators that generate nonzero weights for the qutrit model, equivalently, sequences of $P_1$ and $P_2$ that multiply to the identity.

\begin{table*}
\begin{tabular}{|c|c|c|c|}
\hline
expansion order $q$ & $S_{{\bf{i}}_q}=\mathbb{1}$  & phase & sign of weight\\
\hline
0 & $\mathbb{1}$ &$ 0$& 1\\
1& --  & --&--\\
2&  $P_1 P_2, P_2 P_1$  & $0$& 1\\
3 &  $P_1 P_1 P_1, P_2 P_2 P_2$ & $3\phi$&$\text{sgn}(\cos 3\phi)$\\
4 &   $P_1 P_2 P_1 P_2$ and permutations thereof  & $0$ &1 \\
5 &  $P_1 P_1 P_1 P_1 P_2, P_2 P_2 P_2 P_2 P_1$ and perms. thereof &  $3\phi$ &$\text{sgn}(\cos 3\phi)$\\
\hline
\end{tabular}
\caption{\label{tab:paths}Operator sequences in the lowest expansion orders, the associated weights and signs.}
\end{table*}

It would be very instructive to analyze the onset of the sign problem in this simple model. 
Let $n_1$ ($n_2$) be the number of $P_1$ ($P_2$) operators in a given sequence of operators $S_{{\bf{i}}_q}$. The sequence evaluates to $P_1^{n_1+2n_2}$ as $P_2=P_1^2$. Since the product $S_{{\bf{i}}_q}$ must evaluate to the identity element for the weight to be nonzero and $P_1^{3} = \mathbb{1}$ we find that the condition translates to \hbox{$(n_1+2n_2) \bmod3=0$}, or  \hbox{$n_1 + 2n_2 = 3m$} for some integer $m$. Moreover, the sign of the weight associated with $W_\C$ is 
\begin{eqnarray}\label{eq:sgnCal}
\sgn(W_\C) &=& \sgn \; (\Re D_{(z,S_{{\bf{i}}_q})}) \cdot \sgn(e^{-\beta [E_{z_0}, \ldots, E_{z_q}]}) \nonumber  \\
&=& \sgn \;\Re e^{i (n_1-n_2) \phi } \cdot \sgn(-1)^{n_1+n_2} \nonumber  \\
&=& \sgn \; \Re e^{i \pi ( (n_1-n_2)\frac{\phi}{\pi} + (n_1+n_2) ) } \,.
\end{eqnarray}
For the model to be sign-problem free, the phase of $W_\C$ must be a multiple of $2\pi$, which yields
\begin{eqnarray}\label{eq:sgnCond1}
[ (n_1-n_2)\frac{\phi}{\pi} + (n_1+n_2) ] \bmod 2 = 0 \,.
\end{eqnarray} 
Plugging $n_1 = 3m-2n_2$ into Eq.~(\ref{eq:sgnCond1}) gives
\begin{eqnarray}\label{eq:sgnCond2}
{[ (1+3\frac{\phi}{\pi})(m-n_2)]} \bmod 2 = 0 \,.
\end{eqnarray}
For Eq.~(\ref{eq:sgnCond2}) to hold for any $m$ and $n_2$, we must require that $1+3(\phi/\pi)$ is a multiple of 2. That is, $1+3(\phi/\pi) = 2k$ for some integer $k$. This in turn gives 
$\phi=(2k+1)\pi/3$, meaning that in the range  $\phi \in [0,2\pi)$ there are precisely three values for our qutrit model to have no sign problem. These are $\phi_* \in \{\pi/3, \pi, 5 \pi/3\}$.

\begin{figure}[htp]
\includegraphics[width=0.45\textwidth]{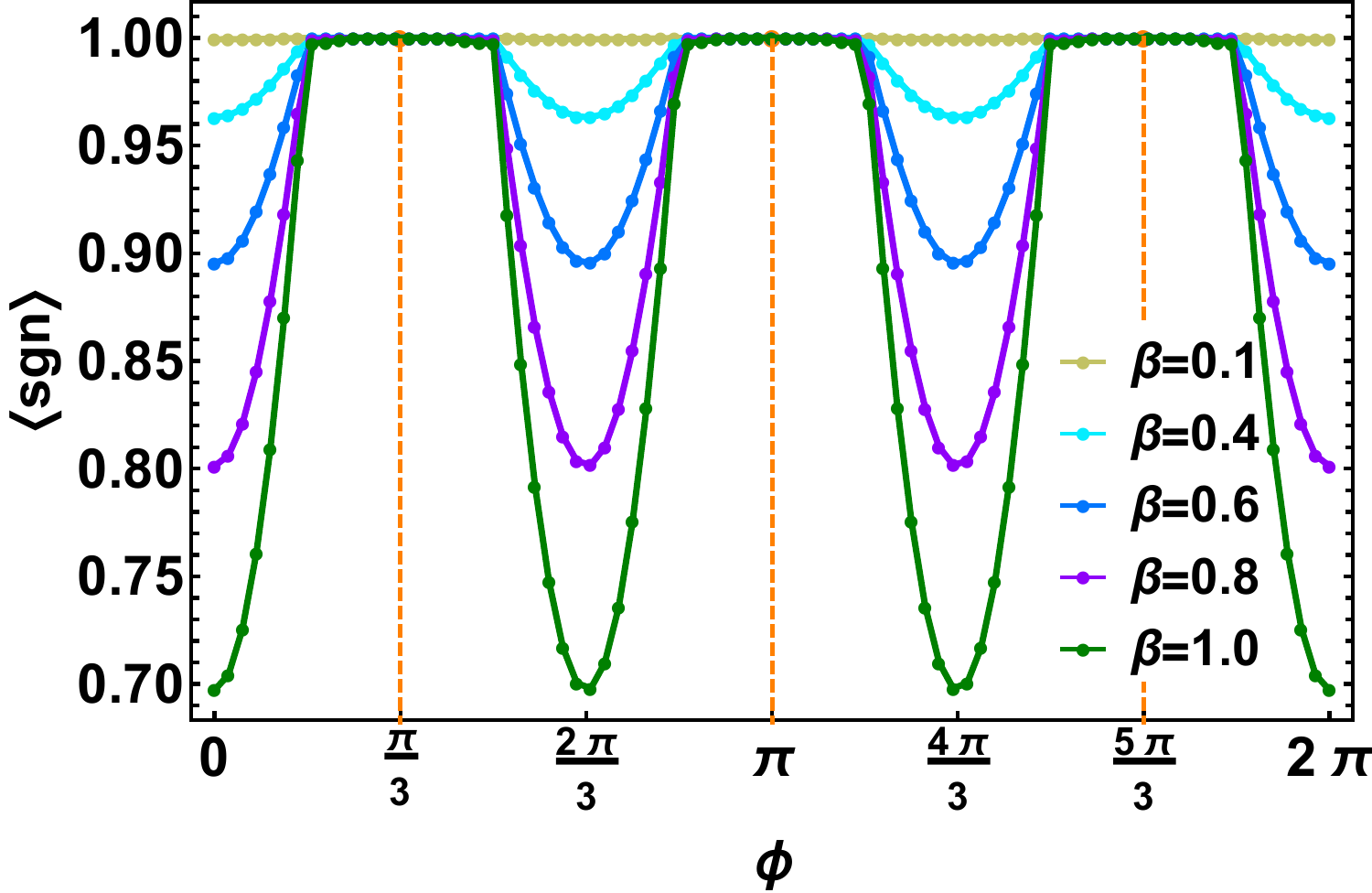}
\includegraphics[width=0.45\textwidth]{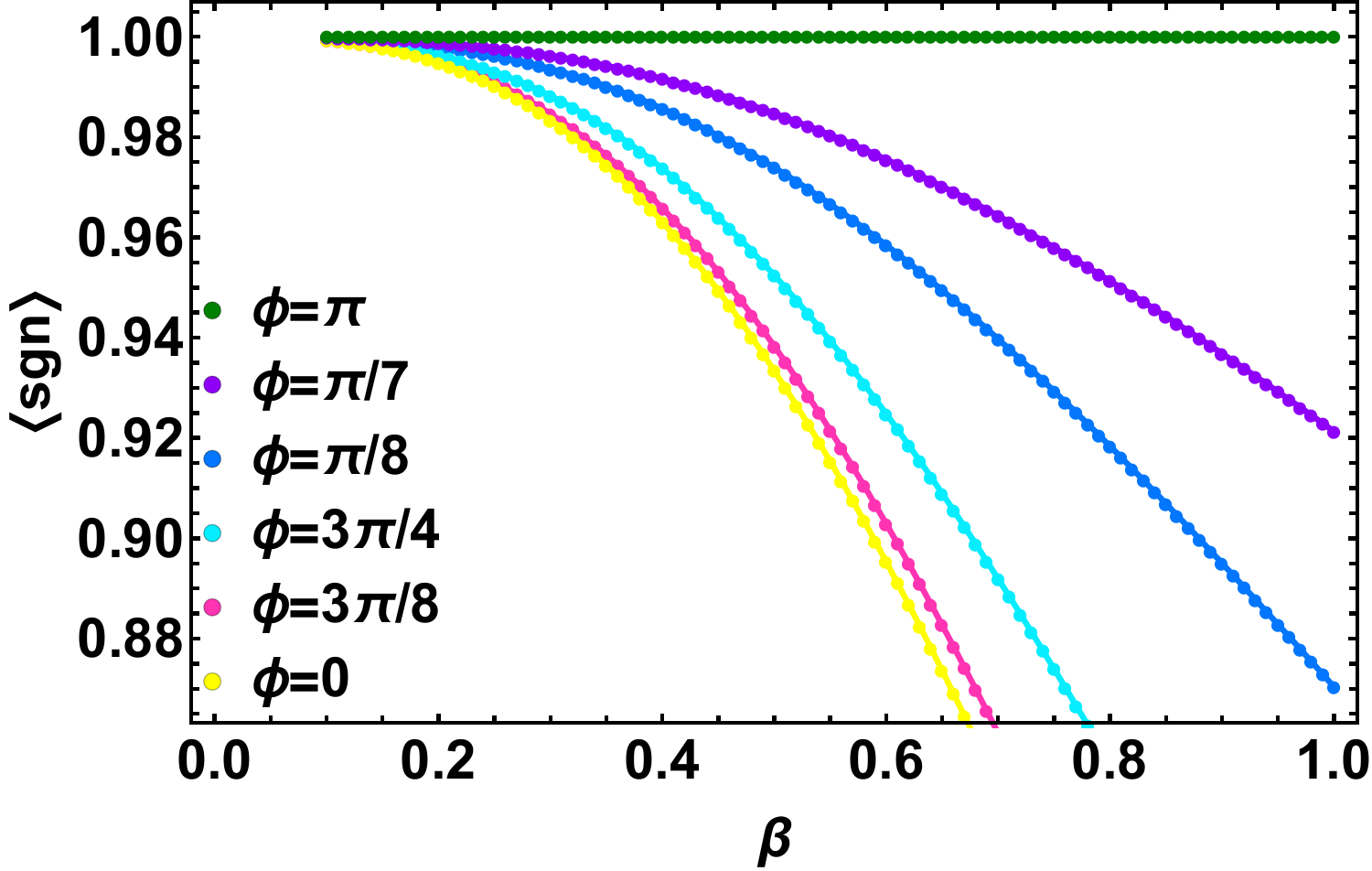}
\small
\caption{\label{fig:signQutrit} Top: The average sign $\langle \text{sgn} \rangle$ as a function of $\phi$ for various values of $\beta$. The severity of the sign depends crucially on the phase $\phi$. 
For three values of $\phi$ (marked by vertical lines) the model is sign-problem free. Bottom: $\langle \text{sgn} \rangle$ as a function of $\beta$ for various values of $\phi$, illustrating the fact that the problem becomes more severe with increasing $\beta$.} 
\end{figure}

We corroborate our analytical derivation by numerical simulations. 
In Fig.~\ref{fig:signQutrit}(top) we examine the behavior of $\langle \text{sgn} \rangle$ as a function of phase $\phi$ for fixed values of $\beta$. 
As is evident from the figure, the severity of the sign problem oscillates periodically between the `worst case' values at $\{0, 2\pi/3,4 \pi/3\}$ to the sign-problem-free values in between. Figure~\ref{fig:signQutrit}(bottom) shows that for a fixed value of phase $\phi$ the sign problem becomes exponentially more severe with inverse-temperature.

The qutrit example analyzed above demonstrates that not only is it possible for a single particle model to have sign problem, but that it is also possible to control the severity of the sign problem by tuning the phase of off-diagonal terms while keeping their magnitudes fixed. The fact that the model is sign-problem-free for $\phi_\star=\pi/3, \pi$ and $5 \pi/3$ very clearly illustrates that non-stoquasticity, i.e., the complex-valuedness of off-diagonal Hamiltonian matrix entries do not necessarily induce a sign problem. Rather, it is the phase acquired along paths that induces it. \\

The phase-dependent sign problem of the single qutrit model above can readily be generalized to the more interesting many-body case, e.g., the Hamiltonian
\beq\label{eq:qutritHamMulti}
H= \sum_j D_0^j + \sum_{\langle i j\rangle} \left( \e^{i \phi} P_1^i P_2^j + \e^{-i \phi} P_1^j P_2^i \right)\,,
\eeq
where $\langle i j\rangle$ denotes summation over the edges of an arbitrary interaction graph with qutrits sitting on its vertices.
Using the same arguments as above, it can be shown that Hamiltonian Eq.~(\ref{eq:qutritHamMulti}) possesses a sign problem for certain values of the angle $\phi$ but not for others.

\subsection{The sign problem and positivity of ground-state amplitudes}

In previous sections we have shown that stoquastic models, i.e., models whose Hamiltonians have only nonpositive off-diagonal entries, are sign-problem-free, while the converse is not necessarily true, that is, non-stoquastic Hamiltonians may or may not possess a sign problem. 

Another property that may be derived from stoquasticity, due to the Perron-Frobenious theorem~\cite{perron}, is that their ground state amplitudes all have the same sign. 
This raises the natural question of whether models whose ground-state amplitudes are all positive can even exhibit a sign problem (or a severe sign problem). In what follows, we provide a simple example that demonstrates that this is indeed the case, namely, that a model may have a severe sign problem despite having a ground-state whose amplitudes are all positive. 

To that aim, we consider the simple Hamiltonian
\bea\label{eq:Hp}
H = -(P + P^3) + \epsilon P^2 \,,
\eea
where $P$ is permutation matrix satisfying $P^4 = \mathbb{1}$ but $P^2 \neq \mathbb{1}$. (To simplify matters we have chosen a Hamiltonian with a vanishing diagonal term however such a term may be easily added on without modifying the end results.)
The smallest matrix that satisfies the above relations is  four-dimensional, e.g.,
\begin{eqnarray}
P = 
\begin{bmatrix}
0 & 0 & 0 & 1 \\
1 & 0 & 0 & 0 \\
0 & 1 & 0 & 0 \\
0 & 0 & 1 & 0
\end{bmatrix}  \,.
\end{eqnarray}

The above Hamiltonian, Eq.~(\ref{eq:Hp}), has three off-diagonal terms \hbox{$\{P,P^2,P^3\}$} whose respective diagonal operators are \hbox{$\{D_1=-\mathbb{1},D_2=\epsilon \mathbb{1},D_3=- \mathbb{1} \}$}. It has a sign problem for all $\epsilon>0$ (in which regime the model is non-stoquastic). This is due to sequences of odd orders (e.g., the $q=3$ sequence $S_{{\bf{i}}_q}= P \cdot P^2 \cdot P$) having negative weights, owing to the positivity of $\epsilon$. 
Nonetheless, inspecting the spectrum of the model we find that it has eigenvalues 
\hbox{$(-2 + \epsilon,-\epsilon,-\epsilon,2 + \epsilon)$} with the respective eigenvectors
\hbox{$\{ 1, 1, 1, 1 \},\{ 0, -1, 0, 1 \},\{ -1 , 0, 1, 0 \}$} and $\{ -1 , 1, -1, 1 \}$ which in turn implies that for $\epsilon<1$, the ground state is $\{1,1,1,1\}$, i.e., it the has all-positive amplitudes.

\begin{figure}[htp]
\includegraphics[width=0.49\textwidth]{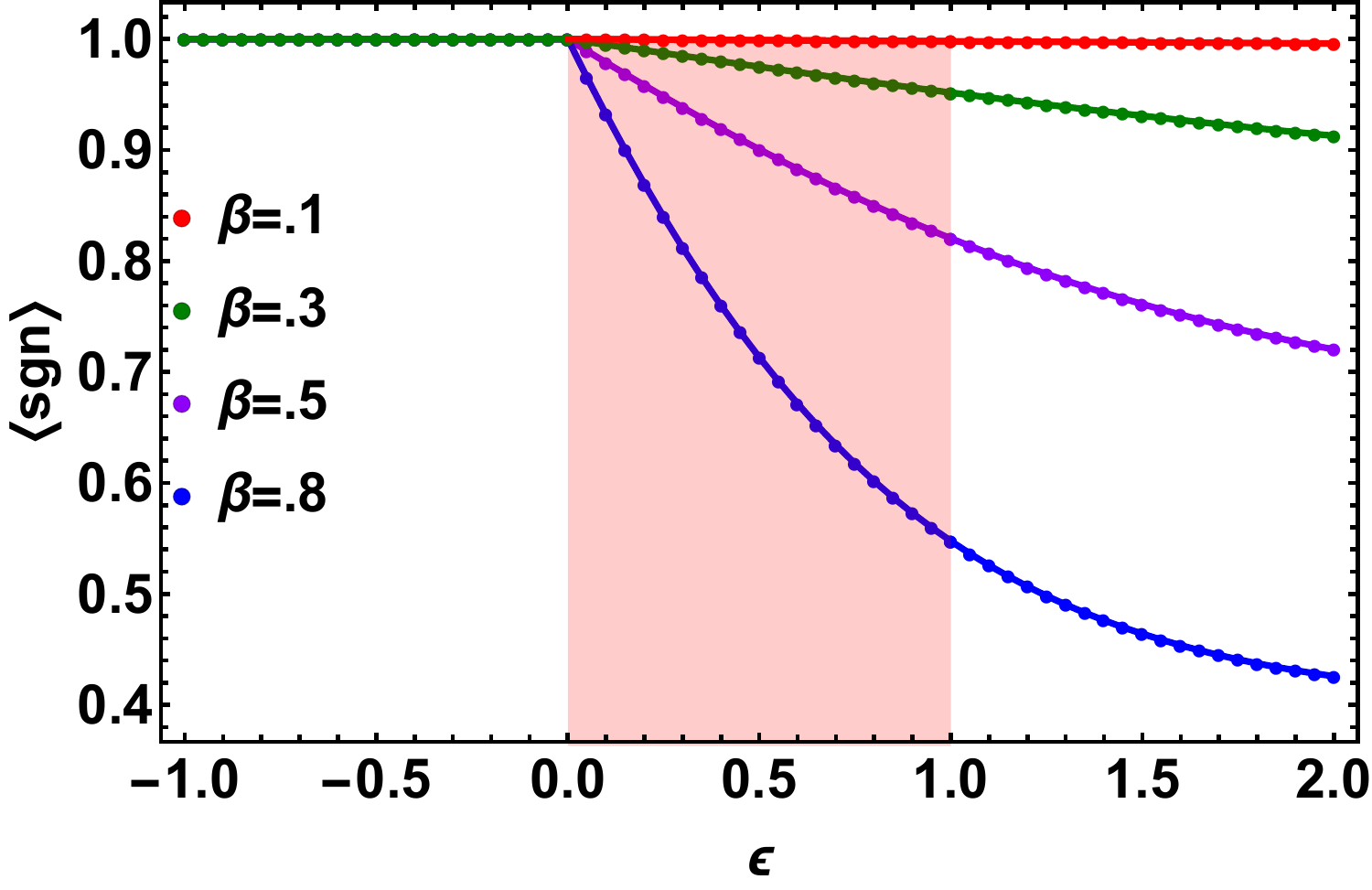}
\small
\caption{\label{fig:PermSgn} The severity of the sign problem $\langle \text{sgn} \rangle$ as a function of $\epsilon$ for the Hamiltonian $H = -(P + P^3) + \epsilon P^2$. The model is sign-problematic for all $\epsilon > 0$. In the region $0 < \epsilon < 1$ all ground state amplitudes have the same sign. } 
\end{figure}

Thus, in the region $0 < \epsilon <1$, the model is sign problematic and has all-positive ground-state amplitudes. In fact $\langle \text{sgn}\rangle$ can be calculated analytically in this case to be 
\bea
\langle \text{sgn}\rangle &=& \frac{\Tr\left[\e^{-\beta H}\right]}{\Tr\left[\e^{-\beta H_\s}\right]} \nonumber \\
&=& \frac{2 e^{2 \beta  (\epsilon +1)}+e^{4 \beta }+1}{e^{2 \beta  \epsilon }+e^{2 \beta 
   (\epsilon +2)}+2 e^{2 \beta }}. \;
\eea
As is depicted in Fig.~\ref{fig:PermSgn} and can be seen from the above expression, the sign problem becomes exponentially more severe as the inverse-temperature $\beta$ increases (regardless of the ground-state sign). 

\section{Conclusions and discussion}

In this study we made an attempt to elucidate the QMC sign problem, one of the most fundamental bottlenecks of many-body physics simulations~\cite{Wiese-PRL-05,marvianLidarHen,klassenMarvian,signProbSandvik}, and alleviate the confusion surrounding certain aspects of the problem. Along the way, we derived a formalism that allowed us to characterize the conditions under which the sign problem emerges and to analyze several core examples that illustrate a number of important observations.

First and foremost, we demonstrated that non-stoquasticity --- the non-positivity or complex-valuedness of the off-diagonal entries of the matrix representation of Hamiltonians --- does not imply the existence of a sign problem. We have shown rather that the emergence of a sign problem is tightly connected to the cumulative phase of products of off-diagonal terms along closed paths in the hypercube of basis states. 

We have also shown that a single spin-$1/2$ particle cannot possess a sign problem but that a single spin-one particle can. We have additionally provided an example illustrating that a physical model may be non-stoquastic and have a severe sign problem despite having all-positive ground-state amplitudes, a property that is usually linked with stoquasticity~\cite{perron}. 

Developing a true understanding of the nature of the QMC sign problem is important in physics, chemistry, the material sciences and well beyond those, and is crucial to the potential resolution of the problem. We therefore hope that our work will provide a useful framework for studying the nature of the sign problem in models of physical relevance beyond those analyzed here and will allow addressing the sign problem in more general settings. 

Another area in which developing a true understanding of the problem is important is the field of quantum computing. This is because quantum simulations~\cite{Childs9456,Reiher201619152}, or simulations of quantum many-body systems on quantum computers, as originally envisioned by Richard Feynman~\cite{Feynman:QC} is one of the most promising applications of near-term quantum devices (as well as the more distant fault-tolerant universal quantum computers). The existence of quantum simulation speedups hinges on the premise that simulating quantum systems is an intractable task for standard computers, and relies (at least in part) on the intractability of the sign problem.
In this context, it is worth noting in particular the common misconception that the resolution of the sign problem would imply in general a polynomial-time equilibration of QMC simulations or that it would somehow provide as a result a resolution to the famous P versus NP problem of computer science~\cite{PNP}. Notwithstanding, a general resolution to the sign problem would imply the existence of efficient classical simulations to quantum algorithms~\cite{cerf1997monte,Howard_2017,hastings2013obstructions}. 

\vspace{.5cm}

\begin{acknowledgements}
The research is based upon work supported by the Office of the Director of National Intelligence (ODNI), Intelligence Advanced
Research Projects Activity (IARPA), via the U.S. Army Research Office
contract W911NF-17-C-0050. This material is based on research sponsored by the Air Force Research laboratory under
agreement number FA8750-18-1-0044. The U.S. Government is authorized to reproduce and distribute
reprints for Governmental purposes notwithstanding any copyright notation thereon."
The views and conclusions contained herein are
those of the authors and should not be interpreted as necessarily
representing the official policies or endorsements, either expressed or
implied, of the ODNI, IARPA, or the U.S. Government. 
\end{acknowledgements}

\bibliography{refs}

\newpage
\pagebreak
\appendix 
\section{Notes on divided differences} \label{sec:DividedDifference}
We provide below a brief summary of the concept of divided differences which is a recursive division process. This method is typically encountered when calculating the coefficients in the interpolation polynomial in the Newton form.

The divided differences~\cite{dd:67,deboor:05} of a function $F(\cdot)$ is defined as
\beq\label{eq:divideddifference2}
F[x_0,\ldots,x_q] \equiv \sum_{j=0}^{q} \frac{F(x_j)}{\prod_{k \neq j}(x_j-x_k)}
\eeq
with respect to the list of real-valued input variables $[x_0,\ldots,x_q]$. The above expression is ill-defined if some of the inputs have repeated values, in which case one must resort to a limiting process. For instance, in the case where $x_0=x_1=\ldots=x_q=x$, the definition of divided differences reduces to: 
\beq
F[x_0,\ldots,x_q] = \frac{F^{(q)}(x)}{q!} \,,
\eeq 
where $F^{(n)}(\cdot)$ stands for the $n$-th derivative of $F(\cdot)$.
Divided differences can alternatively be defined via the recursion relations
\bea\label{eq:ddr}
&&F[x_i,\ldots,x_{i+j}] \\\nonumber
&=& \frac{F[x_{i+1},\ldots , x_{i+j}] - F[x_i,\ldots , x_{i+j-1}]}{x_{i+j}-x_i} \,,
\eea 
with $i\in\{0,\ldots,q-j\},\ j\in\{1,\ldots,q\}$ with the initial conditions
\beq\label{eq:divideddifference3}
F[x_i] = F(x_{i}), \qquad i \in \{ 0,\ldots,q \}  \quad \forall i \,.
\eeq
A function of divided differences can be defined in terms of its Taylor expansion. In the case where $F(x)=\e^{-\beta x}$, we have
\beq
\e^{-\beta [x_0,\ldots,x_q]} = \sum_{n=0}^{\infty} \frac{(-\beta)^n [x_0,\ldots,x_q]^n}{n!} \ . 
\eeq 
\end{document}